\documentclass{article}
\usepackage{spconf,amsmath,graphicx}
\usepackage{epsfig}
\usepackage{url}
\usepackage{subfigure}
\usepackage{color}
\usepackage{changepage}
\usepackage{times}
\usepackage{float}
\usepackage{amssymb}
\usepackage{multirow}
\usepackage{bbding}
\usepackage{booktabs}
\usepackage{algorithm}
\usepackage{algpseudocode}

\title{A multiple attributes image quality database for \\smartphone camera photo quality assessment}
\name{\normalsize Wenhan Zhu$^{\dagger \star}$, Guangtao Zhai$^{\star}$, Zongxi Han$^{\star}$, Xiongkuo Min$^{\star}$, Tao Wang$^{\star}$, Zicheng Zhang$^{\star}$ and Xiaokang Yang$^{\dagger}$}

  \address{ $^{\dagger}$ MoE Key Lab of Artificial Intelligence, AI Institute, Shanghai Jiao Tong University, China \\
      $^{\star}$Institute of Image Communication and Information Processing, Shanghai Jiao Tong University, China}
      
\begin{document}
%

\maketitle
\begin{abstract}
\vspace{-0.1cm}
Smartphone is the superstar product in digital device market and the quality of smartphone camera photos (SCPs) is becoming one of the dominant considerations when consumers purchase smartphones. How to evaluate the quality of smartphone cameras and the taken photos is urgent issue to be solved. To bridge the gap between academic research accomplishment and industrial needs, in this paper, we establish a new Smartphone Camera Photo Quality Database (SCPQD2020) including 1800 images with 120 scenes taken by 15 smartphones. Exposure, color, noise and texture which are four dominant factors influencing the quality of SCP are evaluated in the subjective study, respectively. Ten popular no-reference (NR) image quality assessment (IQA) algorithms are tested and analyzed on our database. Experimental results demonstrate that the current objective models are not suitable for SCPs, and quality metrics having high correlation with human visual perception are highly needed.
\vspace{-0.1cm}
\end{abstract}
\begin{keywords}
Image quality assessment (IQA), Smartphone camera photo, subjective assessment, no-reference (NR) metrics.
\end{keywords}
\vspace{-0.5cm}
\section{Introduction}
\label{sec:intro}
\vspace{-0.3cm}
Smartphone has been one of the most popular digital devices in the past decades, with more than 300 million smartphones sold every quarter in the world wide. Most of the smartphone vendors, such as Apple, Huawei, Samsung, launch their new flagship smartphones every year. People use smartphone cameras to shoot selfie photos, film scenery or events, and record videos of family and friends. The specifications of smartphone camera and the quality of taken photos are major criteria for consumers to select and purchase smartphones. Many smartphone manufacturers also introduce and advertise their smartphones by introducing the strengths and advantages of their smartphone cameras. Currently in the market, several teams and companies, such as DxOMark \cite{DXO}, evaluate the quality of smartphone cameras. However, the scores and rankings of smartphone cameras they announced are subjectively graded by several photographers and experts, which is not easy to reproduce and deploy in practical image processing systems \cite{WangSSIMFR}. Therefore, automatically and reliably evaluating the quality of photos taken by smartphone cameras becomes a urgent need for smartphone manufacturers and confumers.

In the last two decades, image quality assessment (IQA) has been widely researched in the field of image processing and analysis. Plenty of popular image databases have been constructed to study this topic, for instance, LIVE \cite{LIVE}, CSIQ \cite{CSIQ} and TID2013 \cite{TID2013}. Numerous objective no-reference (NR) IQA algorithms have been proposed to predict the perception of human vision without reference image \cite{BRISQUE}. Most of NR metrics are aimed at evaluating the quality of general distorted images based on different feature extraction methodologies, for example, NIQE \cite{NIQE}, which is developed from spatial natural scene statistic (NSS) features, HOSA \cite{HOSA} that is on account of aggregation of high order statistics, BPRI \cite{min2017blind} and BMPRI \cite{min2018blind} that are based on pseudo reference images and NFERM \cite{NFERM} that applies free energy principle to design the algorithm. In addition, there are some objective models proposed to estimate specific distortions, such as blur \cite{CPBD,FISH} and contrast \cite{fang2014no}. Also, some algorithms are designed for different types of images, such as, screen content images \cite{gu2016saliency} and 360-degree images \cite{yu2015framework}.

However, most of these objective IQA methods are developed to assess the overall perceived quality of the image degraded by various simulated distortions, which rarely exist in photos taken by the modern smartphone cameras. The factors that primarily influence the quality of smartphone camera photos can be classified into some specific properties, such as exposure, color, noise and texture. Thus, the above-mentioned models may be invalid for smartphone camera quality assessment, while objective evaluation methods specifically designed for the purpose of smartphone camera quality assessment are relatively rare. The evaluation for smartphone camera photo (SCP) is still in a lack in the IQA field.

\begin{figure*}[t]
\centering
\vspace{-0.1cm}
\subfigure[] {\label{fige11}\includegraphics[width=0.23\textwidth]{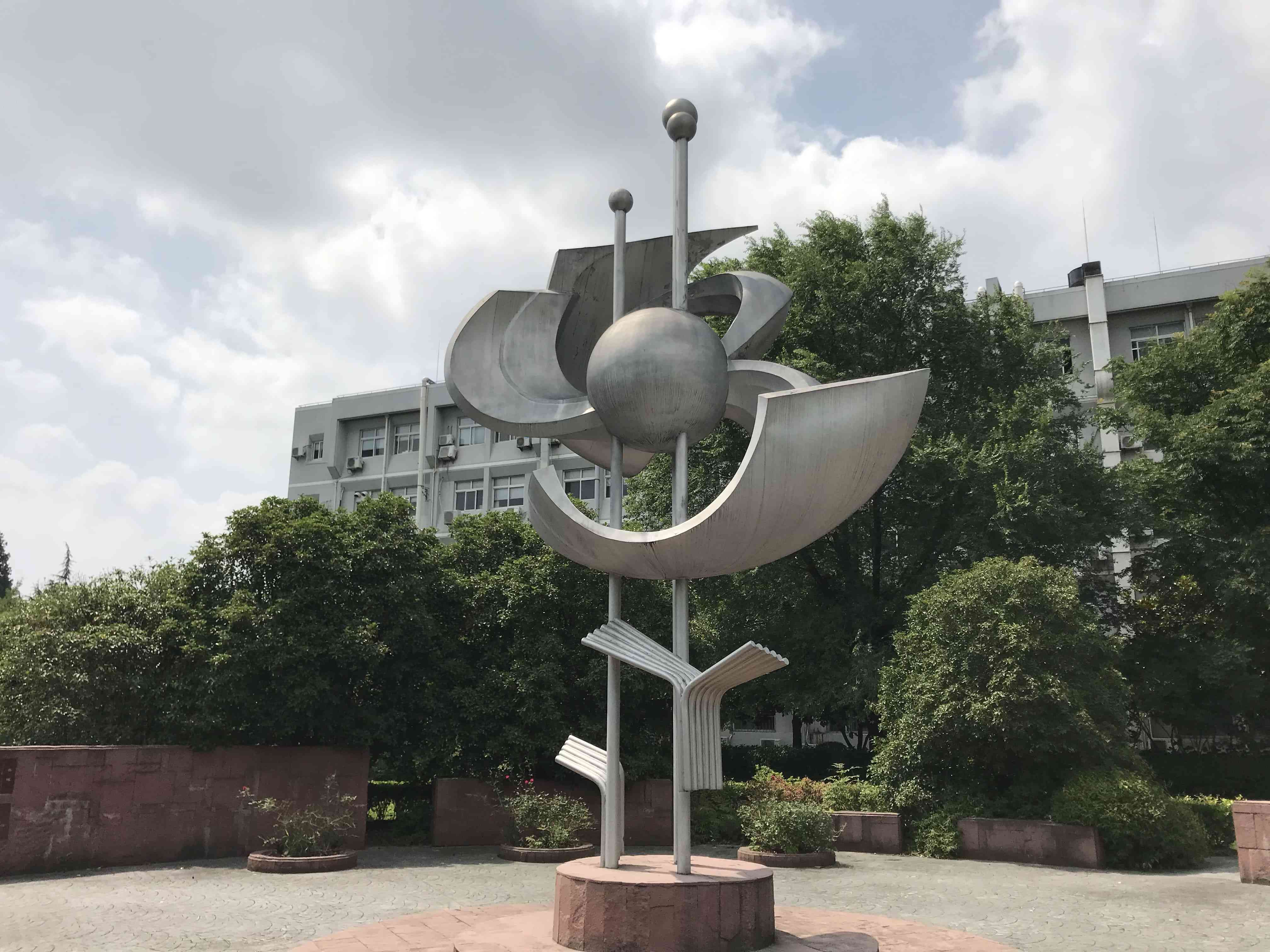}}
\subfigure[] {\label{fige12}\includegraphics[width=0.23\textwidth]{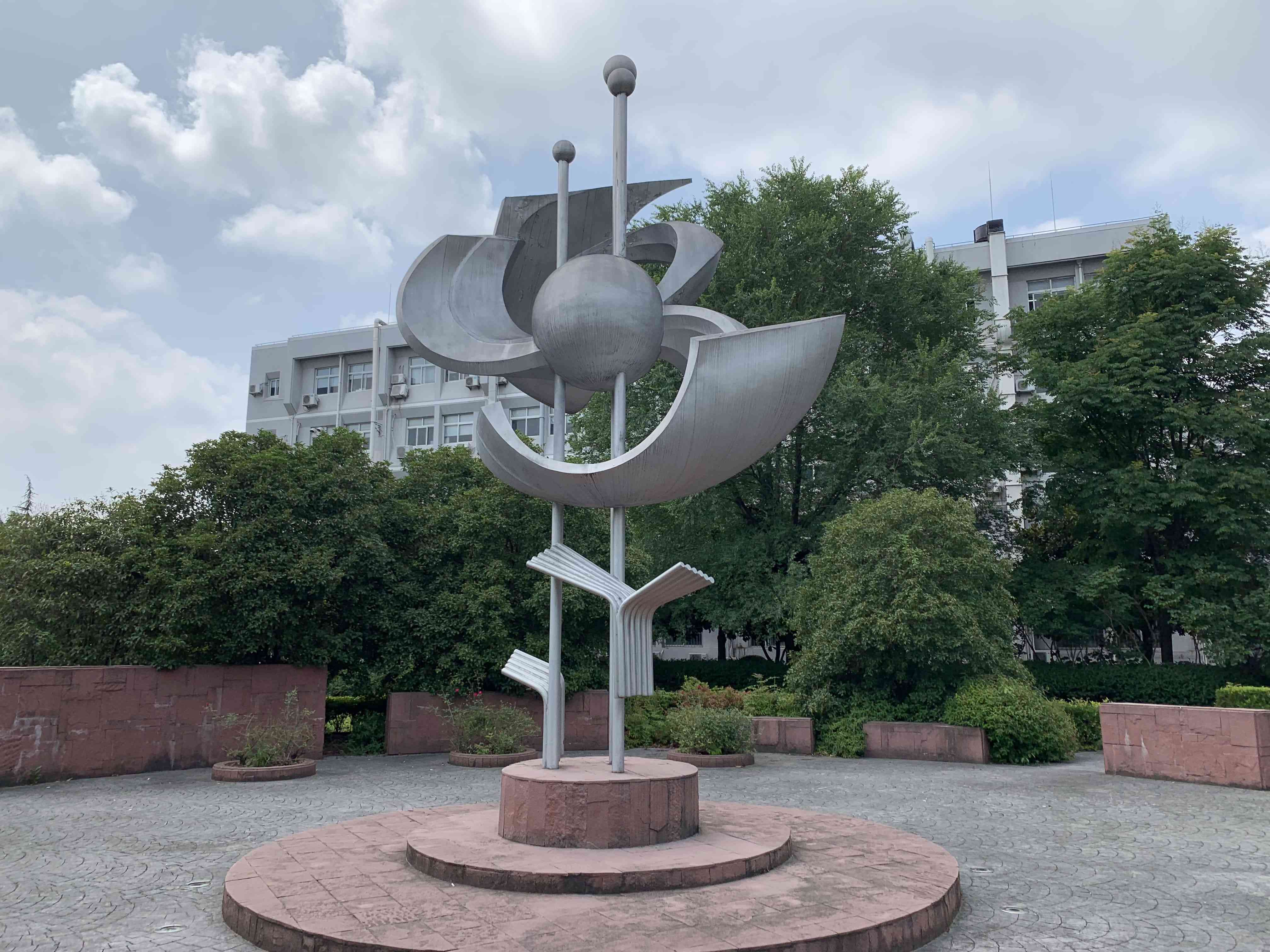}}
\subfigure[] {\label{fige13}\includegraphics[width=0.23\textwidth]{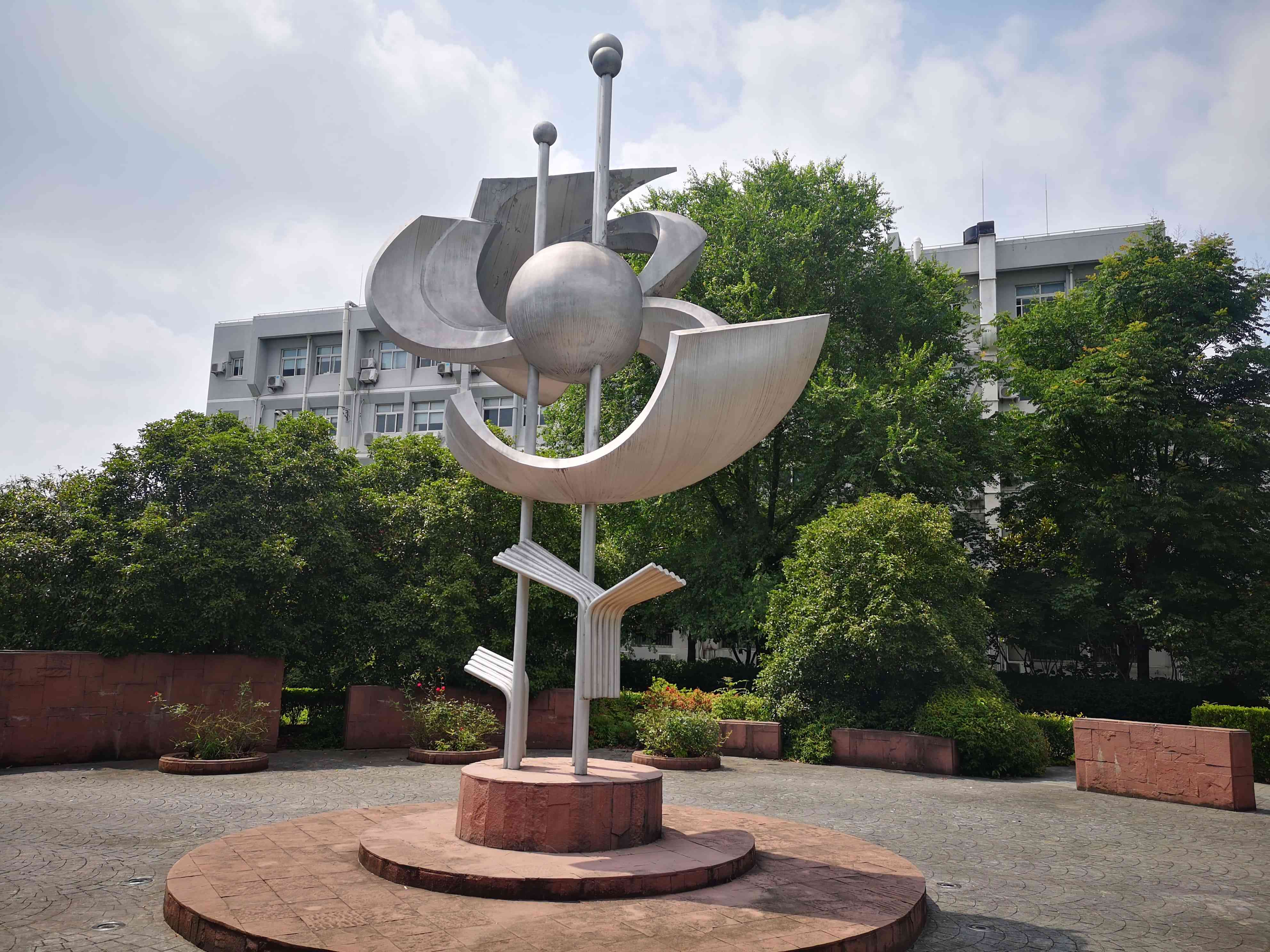}}
\subfigure[] {\label{fige14}\includegraphics[width=0.23\textwidth]{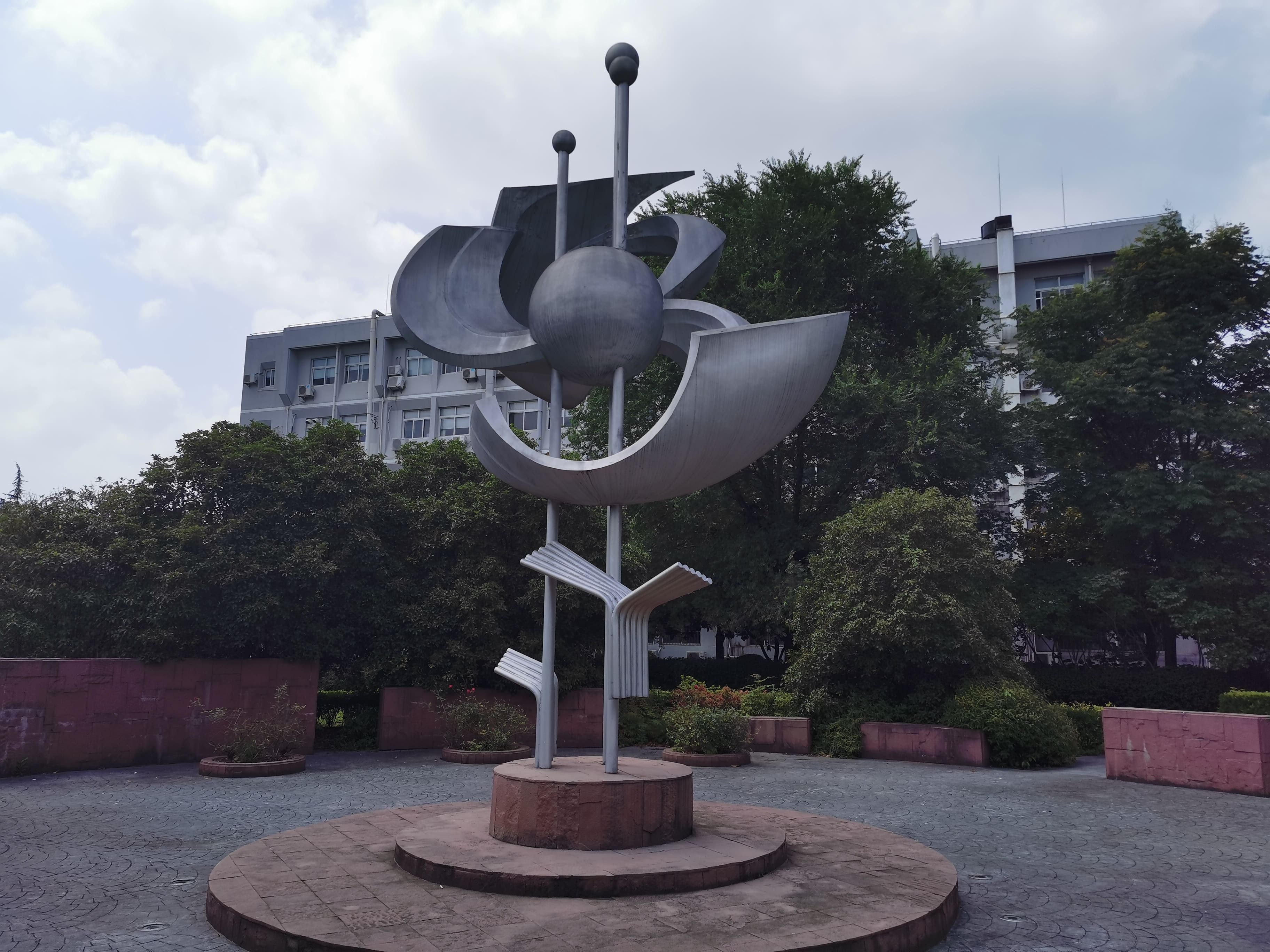}}\\
  \vspace{-0.2cm}
\subfigure[] {\label{fige15}\includegraphics[width=0.23\textwidth]{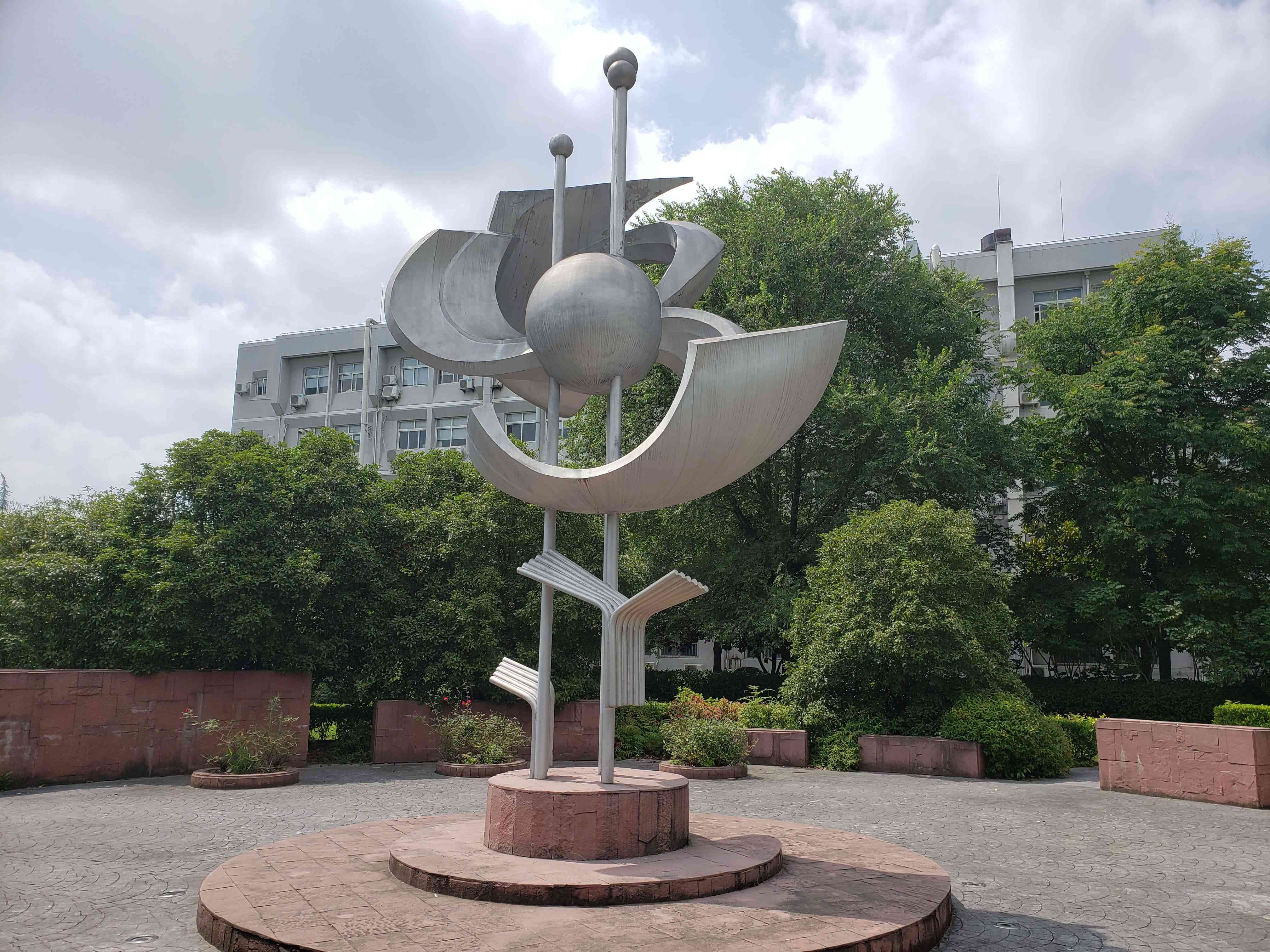}}
\subfigure[] {\label{fige16}\includegraphics[width=0.23\textwidth]{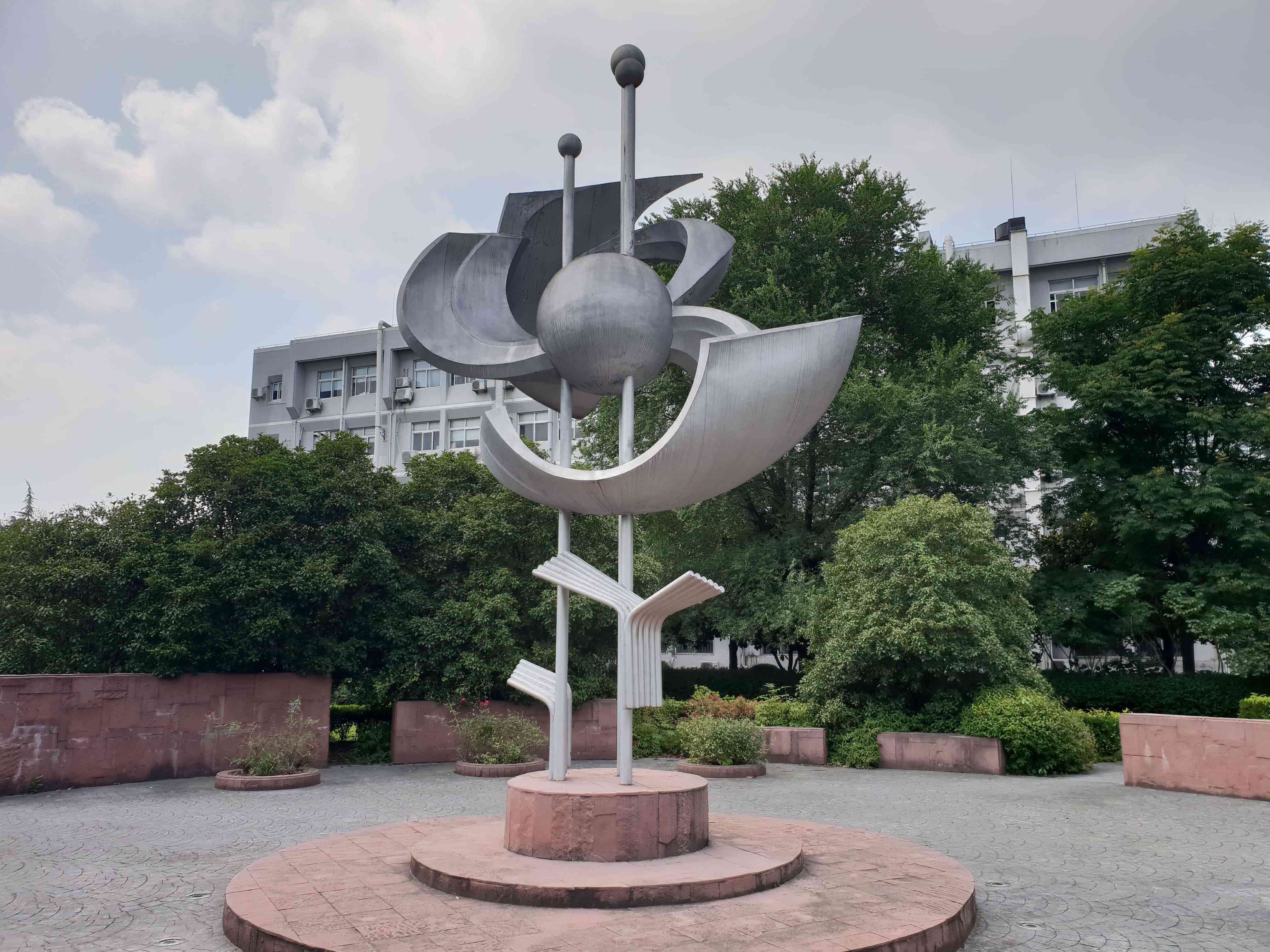}}
\subfigure[] {\label{fige17}\includegraphics[width=0.23\textwidth]{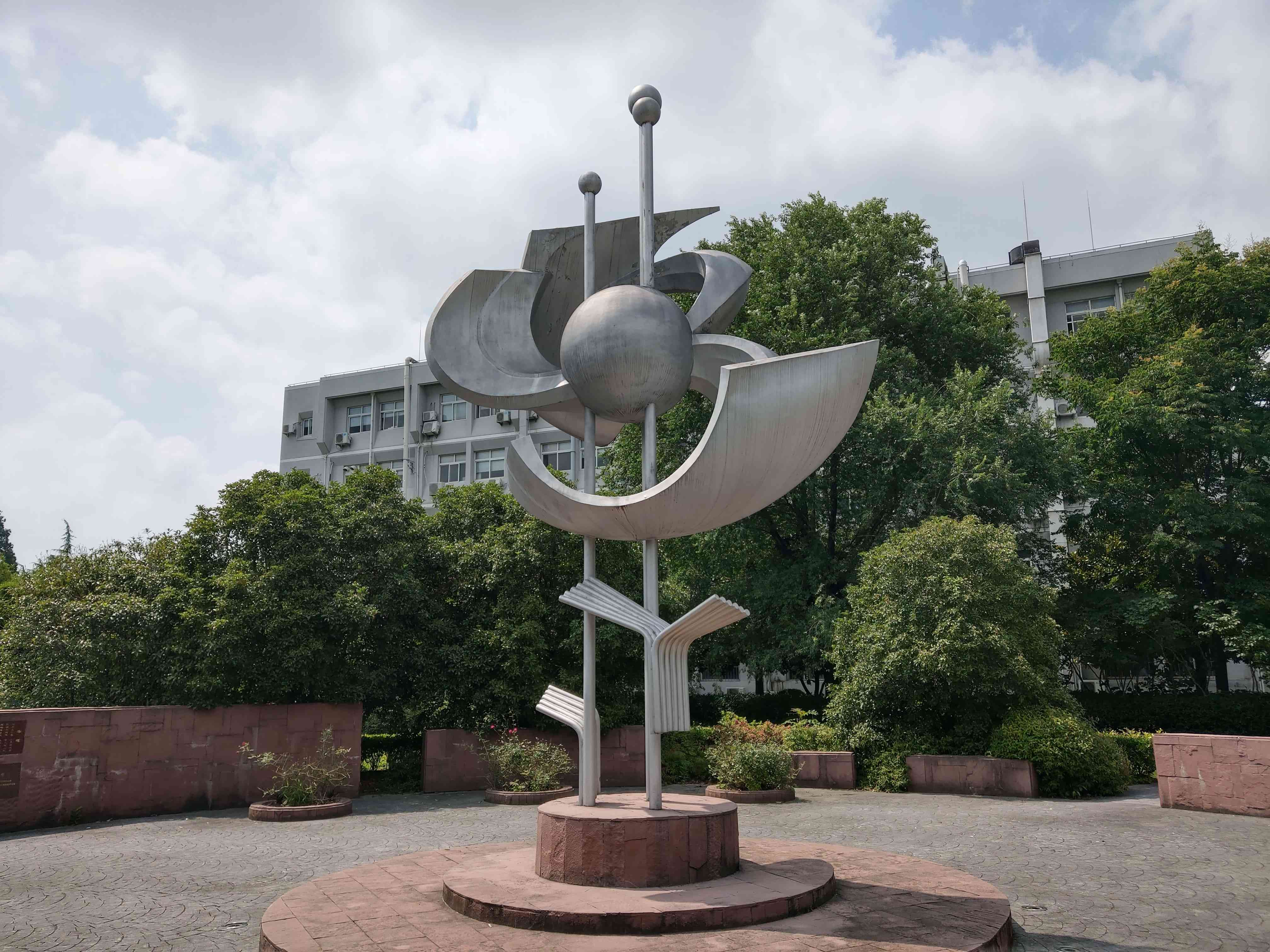}}
\subfigure[] {\label{fige18}\includegraphics[width=0.23\textwidth]{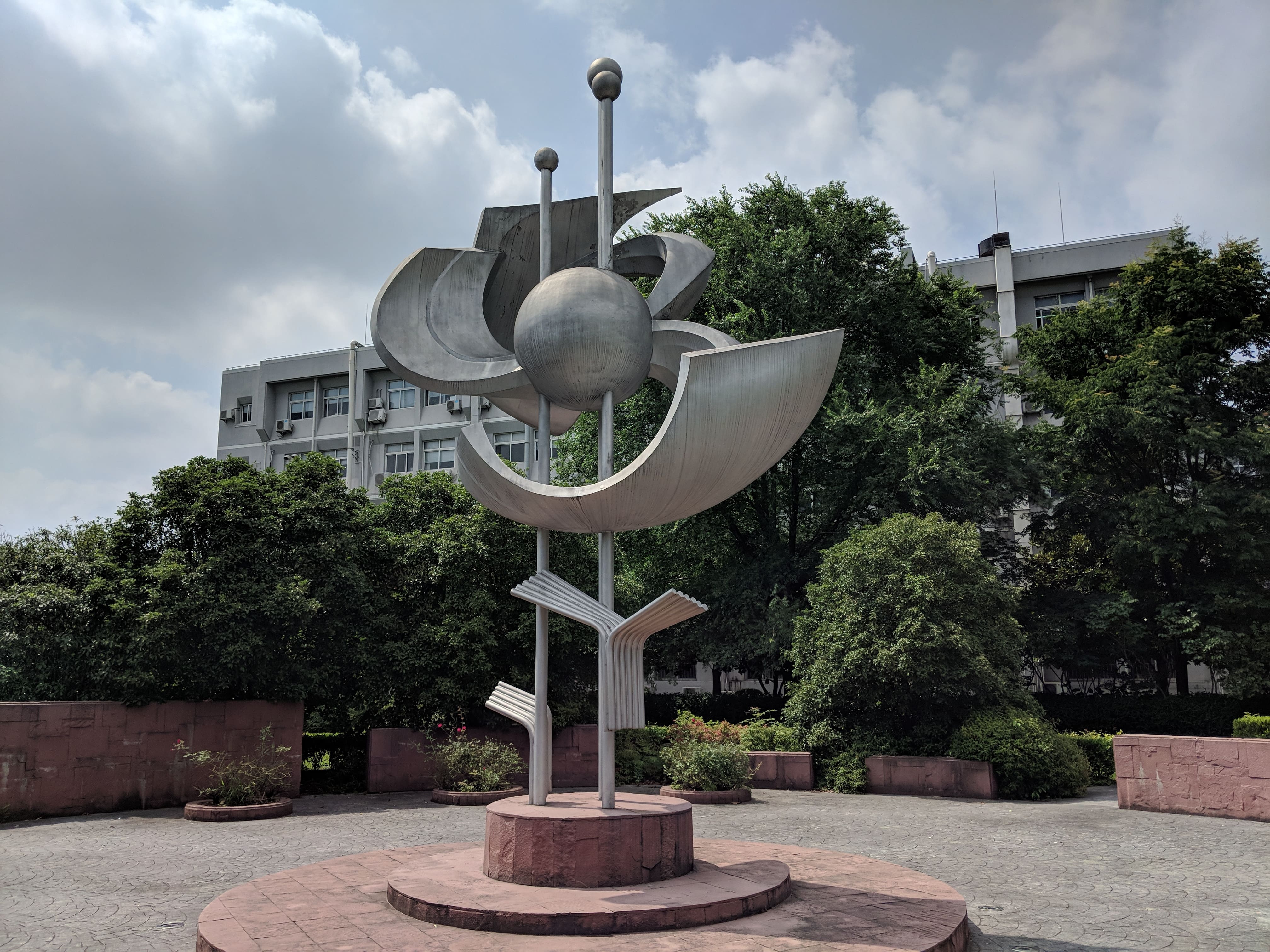}}
  \vspace{-0.2cm}
\caption{Samples of the SCPs taken by eight different smartphones of one scene in SCPQD2020. (a) iPhone 7, (b) iPhone XS max, (c) Huawei P20, (d) Huawei P30 Pro, (e) Samsung Note9, (f) Samsung S8+, (g) HTC U12, (h) Pixel2. }\label{example}
\vspace{-0.5cm}
\end{figure*}

So, in this paper, we establish a new smartphone camera photo quality database for promoting the study of smartphone camera photo quality assessment. Specifically, we first construct a Smartphone Camera Photo Quality Database (SCPQD2020), which consists of 1800 SCPs taken from 120 scenarios using 15 different smartphones. Next, we adopt the purpose-made scoring interface and evaluation criteria to conduct the subjective experiment. Four ratings scored from different aspects, including exposure, color, noise and texture for each SCP are computed via the subjective experiment. After that, we compare ten successful NR IQA metrics on this database. Results demonstrate that the current IQA models do not work well in assessing the quality of SCPs, and quality measures having high correlation with human visual perception are highly needed. A part of this database is already released in the IEEE ICME2020 Grand Challenge \cite{qa4}.


The remainder of this paper is organized as follows. The detailed subjective assessment methodologies for SCPs are introduced in Section 2. Comparisons and evaluations of some objective quality metrics on the SCPQD2020 database are presented in Section 3. Finally, some conclusive remarks are given in Section 4.


\vspace{-0.4cm}
\section{Subjective Quality Assessment}
\vspace{-0.3cm}
\label{sec:sqa}
To investigate quality assessment of SCPs, a targeted image database is constructed which includes SCPs with different smartphones and different scenes. We gather the subjective scores for each SCP from four aspects in the form of MOS.
\vspace{-0.4cm}
\subsection{Image Materials}
\vspace{-0.1cm}
The image database is composed of 1800 photos taken from 120 scenes using 15 smartphones. The goal of this database is to evaluate the photo shooting performances of smartphone cameras designed for ordinary consumers. Therefore, we restore all smartphones to the factory settings and shoot the photos in the default mode. In addition, we ensure that all images for the same scene have the same mainbody content and the same illumination intensity as far as possible. Also, we attempt to avoid the mobile objects, such as walking pedestrians and moving cars, appearing in the center of photos.

These 15 smartphones cover a wide price range and different manufacturers, which are iPhone 7, iPhone X, iPhone XS max, Huawei P20, Huawei P30 Pro, Huawei Mate20 Pro, Samsung S8+, Samsung S9+, Samsung S10+, Samsung Note9, Mi9, HTC U11, HTC U12, Pixel2 and Pixel3. We maintain the original resolutions of these 15 smartphones: eleven of 4032$\times$3024, two of 3648$\times$2736, one of 3968$\times$2976 and one of 4000$\times$3000. For the purpose of ensuring the consistency of the color gamut of all images, we employ the sRGB color gamut. iMazing HEIC Converter \cite{iMazing} is applied for iPhones to convert the Display P3 color gamut to the sRGB color gamut. Our database includes various challenge scenes, e.g. high dynamic scenes, backlight scenes, night scenes, colorful scenes, portrait scenes and distant scenes. Fig. \ref{example} illustrates samples of the SCPs of one scene shooted by eight different smartphones.

\vspace{-0.4cm}
\subsection{Subjective Experiment Methodology}
\begin{figure}
  \centering
  \includegraphics[width=0.45\textwidth]{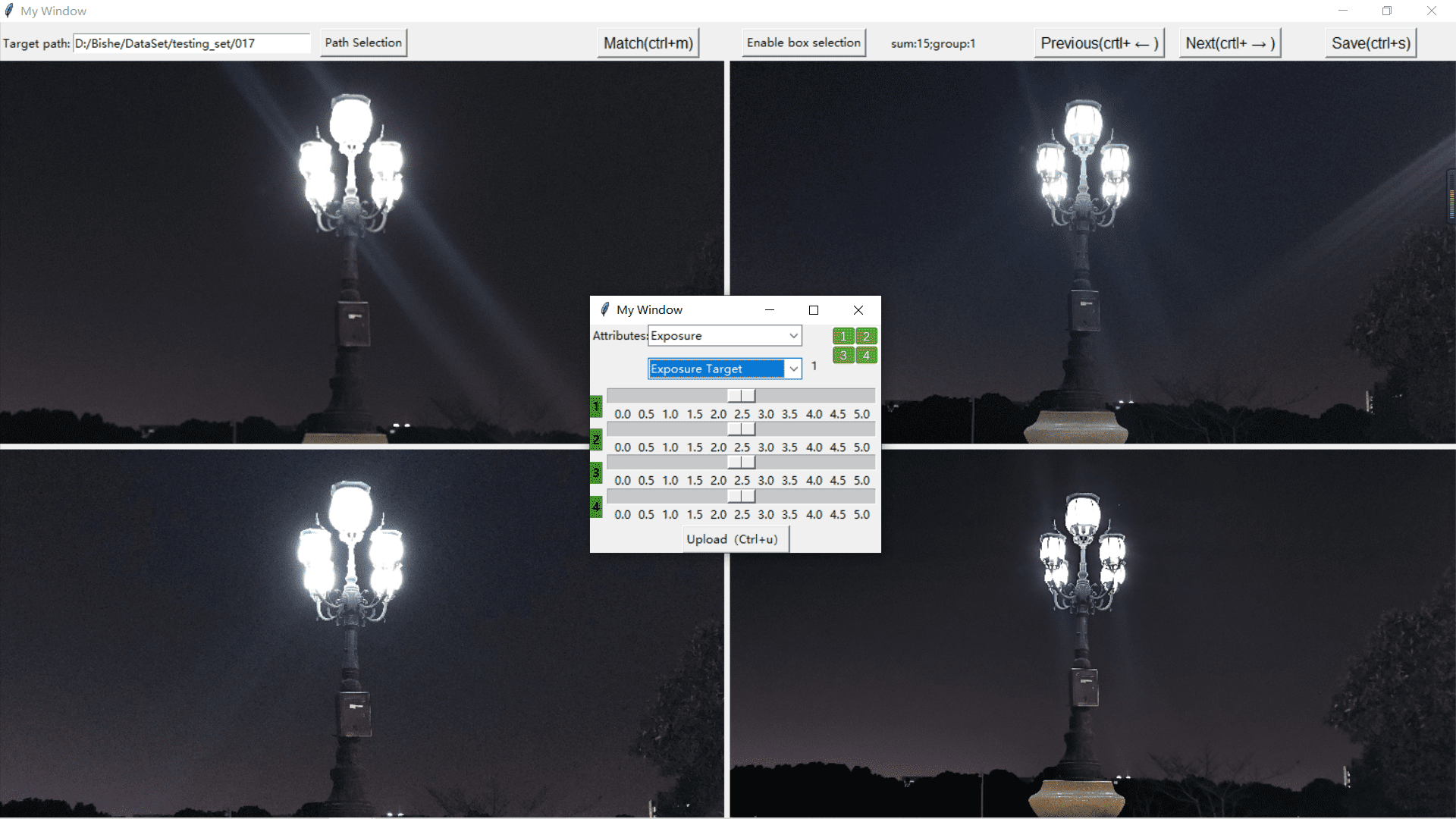}
  \vspace{-0.2cm}
  \caption{Screenshot of the interface of subjective experiments.}\label{jiemian}
  \vspace{-0.3cm}
\end{figure}

\vspace{-0.2cm}
To collect ground truth quality ratings of these SCPs, we conduct the subjective experiment on our database. We develop a user interface using Tkinter in Python to show the photos and gather scores as illustrated in Fig. \ref{jiemian}. In our graphical interface, four photos for a scene taken by different smartphones will appear at a time. Under a zoom number degree of no more than 100\%, the user can select the area of interest (AOI) on any one of four images to zoom in. After determining the AOI of the scene, the same AOI in these four images will be aligned by pressing the ``Match'' bottom. The user can select the attributes, which are exposure, color, noise and texture in our experiment, to score their perceptual ratings.

\begin{figure}
  \centering
  \includegraphics[width=0.45\textwidth]{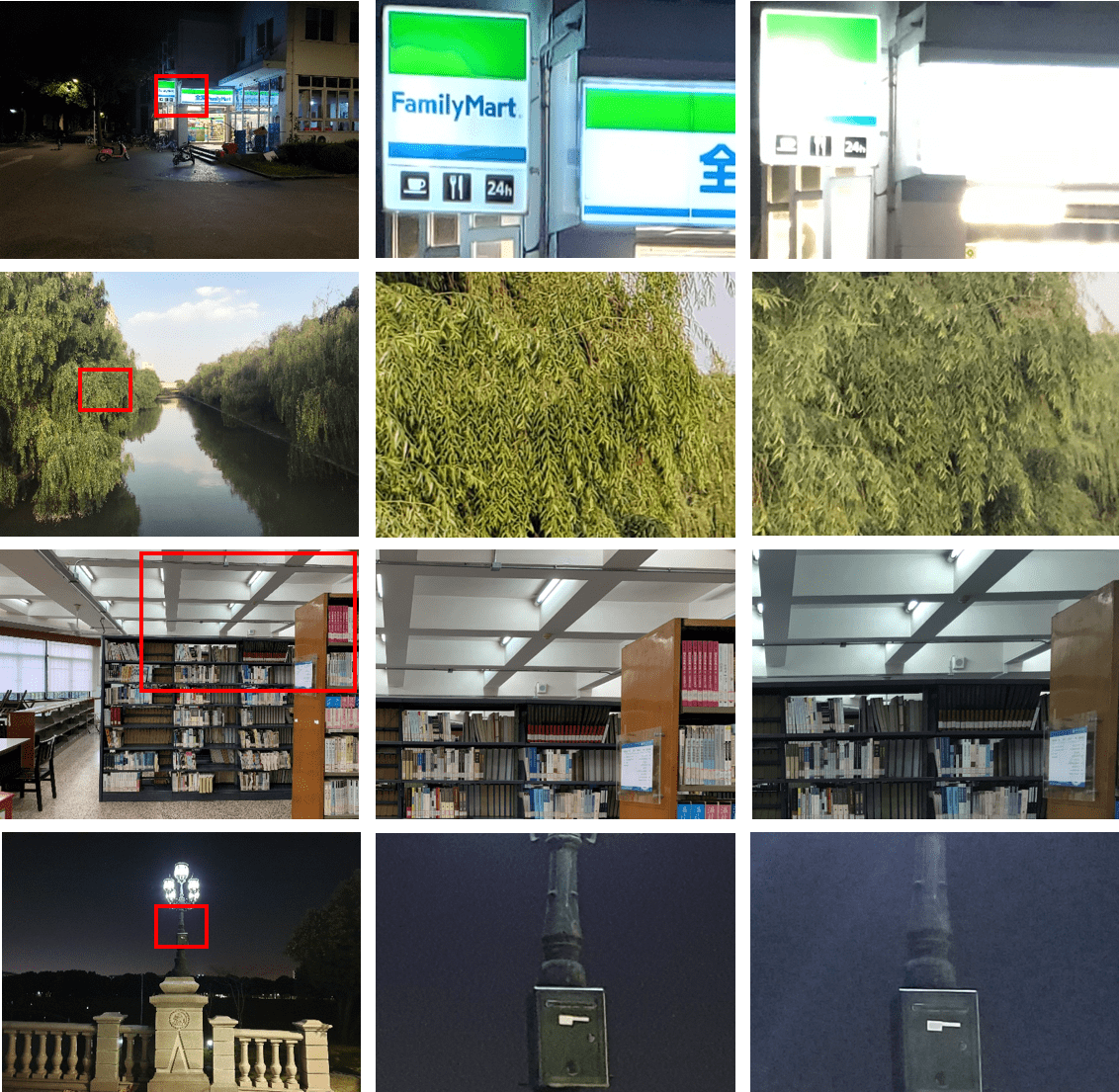}
  \caption{The examples of evaluation criteria in four attributes. From the first row to the fourth row, each row represents the factor of exposure, color, noise and texture, respectively. The first column stands for original photos with AOI. The second column is the regions with high quality in the corresponding attribute. The third column is the regions with low quality in the corresponding attribute.}\label{fourbz}
  \vspace{-0.5cm}
\end{figure}

Unlike other subjective experiments conducted on the traditional image quality assessment, we consider the quality of image in four attributes instead of the overall perception quality. Therefore, we consult the experts of the photography and refer to ITU-R REC. BT.500-13 \cite{assembly2012methodology} to come up with our own standards for evaluation. The quality scale scored from 1 to 5, with a minimum interval of 0.1. Table \ref{score} lists the details of the subjective evaluation criterion of this scale. In addition, we set the evaluation criteria for grading the four attributes (exposure, color, noise and texture) respectively. The details of evaluation criteria are listed as follow and the examples of evaluation criteria are illustrated in Fig. \ref{fourbz}.

    \begin{table}\footnotesize
      \centering
      \caption{The subjective evaluation criterion of quality scale for SCP.}\label{score}
      \begin{tabular}{cccc}
          \toprule[.15em]
       \textbf{Score} & \textbf{Quality} & \textbf{Standards} \\

             \midrule[.15em]
        5 & Excellent & Excellent, pleasant \\
        4 & Good & Fairly clear, slightly distortion \\
        3 & Fair & Perceptible, has minor flaw but passing\\
        2 & Poor & Slightly annoying, has obvious defect\\
        1 & Bad & Quite annoying, unacceptable \\
        \bottomrule[.15em]
      \end{tabular}
\vspace{-0.3cm}
    \end{table}

\begin{itemize}
\vspace{-0.3cm}
  \item[-] \textbf{Exposure:} Observing a region that has both bright part and dark part, the subjects need to determine whether or not there is overexposure and evaluate the degree of overexposure (Especially in night scenes, overexposure may cause losing of details).
      \vspace{-0.3cm}
  \item[-] \textbf{Color:} Looking around the whole photo, the viewers should pay close attention to the problem about color cast and white balance.
  \vspace{-0.3cm}
  \item[-] \textbf{Noise:} Viewing the region that the content is simple and the color is monotonous, such as sky, wall of pure color and dark area, the subjects are required to assess the degree of luminance noise.
      \vspace{-0.3cm}
  \item[-] \textbf{Texture:} Examining the area that has abundant high frequency information, such as leafs, grass and buildings in the distance, the subjects are requested to consider the blurriness, sharpness and reality of this area.
        \vspace{-0.2cm}
\end{itemize}

  \vspace{-0.2cm}

In experiment stage, we employ EIZO RX440 LED displayer with 2560$\times$1600 resolution to show the photos. The viewing distance is set to 1-1.5 times the display height (76 cm). The illuminance of the experimental environment is kept low for comfortable and distinct view. Since this experiment requires knowledge of the field of photos and careful observation, we recruit three experts in photography to assess the quality of SCPs from four attributes. Graders will take a break when they rate for more than half an hour. After the subjective experiment, we gather scores given by all observers. Then, we average the scores of subjects and obtain the final mean opinion score (MOS) of each SCP. Each image's MOS is presented as:

  \begin{equation}\label{MOS}
   MOS^{(p)}_{j}=\frac{1}{N_{i}}\sum^{N}_{i=1}u_{ij}^{(p)}
 \end{equation}
 where $N_{i}$ is the number of valid subjects and $u_{ij}^{(p)}$ is the score of image $j$ in attribute $p$ assigned by the $i$-th subject.
\vspace{-0.3cm}
\section{Comparison of Objective Quality Assessment Models and Discussion}
\vspace{-0.3cm}
%
\begin{table*}\small
  \centering
  \renewcommand{\arraystretch}{1}

  \caption{Performance comparison of the proposed metrics and other IQA methods on SCPQD2020 from four attributes.}\label{results}
  \begin{tabular}{ccccccccccccc}
 \toprule[.15em]
  Attributes & Metrics & BRISQUE & GMLF & HOSA & NIQE & IL-NIQE & LPSI & NFERM & CPBD & BPRI & BMPRI  \\
   \midrule[.15em]
  \multirow{3}{*}{Exposure}  & SRCC & 0.0766 & 0.0542 & 0.0928 & 0.0314 & 0.0367 & 0.1046 & 0.0618 & 0.0590 & 0.1643 & 0.0675  \\
                             & PLCC & 0.1536 & 0.0368 & 0.1779 & 0.1231 & 0.0251 & 0.1815 & 0.1029 & 0.1739 & 0.2021 & 0.1459  \\
                             & RMSE & 10.733 & 23.619 & 6.2960 & 0.8402 & 6.8702 & 0.0171 & 0.1029 & 0.0981 & 0.0118 & 6.9810  \\
\hline
  \multirow{3}{*}{Color}     & SRCC & 0.0759 & 0.0352 & 0.0400 & 0.0143 & 0.0399  & 0.1075 & 0.0405 & 0.0332 & 0.1530 & 0.0755 \\
                             & PLCC & 0.1128 & 0.0442 & 0.1304 & 0.0844 & 0.0438  & 0.1207 & 0.0321 & 0.1103 & 0.1898 & 0.0900 \\
                             & RMSE & 10.793 & 23.577 & 6.3434 & 0.8437 & 6.8625  & 0.0172 & 0.1275 & 0.0990 & 0.0118 & 7.0278 \\
\hline
  \multirow{3}{*}{Noise}     & SRCC & 0.0603 & 0.1569 & 0.1176 & 0.0383 & 0.0539  & 0.0286 & 0.0380 & 0.0228 & 0.2259 & 0.0737 \\
                             & PLCC & 0.1736 & 0.1963 & 0.1612 & 0.0997 & 0.0388  & 0.1164 & 0.0751 & 0.1298 & 0.2505 & 0.1210 \\
                             & RMSE & 10.697 & 23.194 & 6.3144 & 0.8425 & 6.8671  & 0.0172 & 0.1276 & 0.0988 & 0.0149 & 7.0047 \\
\hline
\multirow{3}{*}{Texture}     & SRCC & 0.0948 & 0.0776 & 0.1354 & 0.0993 & 0.0431  & 0.0950 & 0.0813 & 0.0993 & 0.1857 & 0.1023 \\
                             & PLCC & 0.1630 & 0.0960 & 0.2148 & 0.2016 & 0.0371  & 0.2053 & 0.1269 & 0.2199 & 0.2522 & 0.2147 \\
                             & RMSE & 10.716 & 23.545 & 6.2487 & 0.8293 & 6.8676  & 0.0170 & 0.1269 & 0.0972 & 0.0149 & 6.8920 \\

  \bottomrule[.15em]
  \end{tabular}
  \vspace{-0.3cm}
\end{table*}
IQA algorithms for natural image have achieved a remarkable progress over the past decades. Plenty of successful IQA metrics have been provided to automatically predict the perceptual quality of distorted images. However, traditional quality assessment always focus on the overall perception quality of images, while the quality of SCP consider four attributes. Especially, the factors of color and exposure are rarely researched in traditional quality assessment. Thus, the accuracy of existing metrics on evaluating the quality of SCPs needs to be measured and compared.

Since the task of smartphone camera IQA has no reference image, we only select NR IQA algorithms to test. Here, we implement 10 successful NR IQA models which achieve good and reliable performance on traditional image databases. These IQA models are Blind/Referenceless Image Spatial Quality Evaluator (BRISQUE) \cite{BRISQUE}, Gradient Magnitude and Laplacian Features (GMLF) \cite{GMLF}, High Order Statistics Aggregation (HOSA) \cite{HOSA}, Natural Image Quality Evaluator (NIQE) \cite{NIQE}, Integrated Local NIQE (IL-NIQE) \cite{ILNIQE}, Local Pattern Statistics Index (LPSI) \cite{LPSI}, NR Free Energy Based Robust Metric (NFERM) \cite{NFERM}, CPBD \cite{CPBD}, Blind Pseudo Reference Image based (BPRI) quality metric \cite{min2017blind} and Blind Multiple Pseudo Reference Image based (BMPRI) metric \cite{min2018blind}, respectively.

Before calculating performance, we apply a five parameter \{$\beta_{1},\beta_{2},\beta_{3},\beta_{4},\beta_{5}$\} monotonic logistic function to map the scores predicted by objective quality models:
\begin{equation}\label{five}
  Y(x)=\beta_{1}(0.5-\frac{1}{1+e^{\beta_{2}(x-\beta_{3})}})+\beta_{4}x+\beta_{5}
\end{equation}
where $x$ and $Y$ are the objective score and mapped score. Then, we employ three commonly used indices to evaluate the performances of these objective IQA metrics, which are Spearman Rank-Order Correlation Coefficient (SRCC), Pearson Linear Correlation Coefficient (PLCC) and Root Mean Squared Error (RMSE) respectively. To specify, an excellent IQA method is expected to acquire the value close to 1 in SRCC and PLCC, while the value near 0 in RMSE. The performance results of these ten models are listed in Table \ref{results}.

From Table \ref{results}, we find that all of general-purpose NR IQA methods are not of high performance for evaluating quality of SIs, and the majority of their PLCC and SRCC values are lower than 0.2. BPRI, which can adjust algorithm parameters according to different distortion, has the best performance among all methods, while its SRCC and PLCC values do not exceed 0.3. It means that these NR IQA algorithms are noneffective for SCPs.

Considering the ineffectiveness of the above-mentioned NR IQA models, we attempt to analyze the reasons of noneffectiveness for enlightening the design of objective algorithms for SCPs further. Traditional image databases only provide the MOSs of overall perception qualities of distorted images. Whereas, the scores of qualities of SCPs are graded in four attributes: exposure, color, noise and texture. Especially, the factors of exposure and color are rarely researched in the IQA field. For the quality of SCP, the models should be individually designed from different attributes. For instance, the saliency map, luminance histgram and sharpness may be synthesized together for the attribute of exposure; the distribution of color gamut may be an important influence factor for the quality of color.

In addition, traditional distorted images in common databases usually only have single or several distortions artificially added to the reference images. The distortions on images are obvious. Unlike traditional distorted images, the qualities of SCPs are close to the qualities of traditional reference images, even better than them. The resolutions of SCPs are very high. Subjects are required to zoom in to specific areas for observing the distortions. Therefore, how to automatically select the regions for evaluating SCPs is a significant task. The gradient map, saliency map and depth map may be potential directions of creating the region selection algorithms. Thus, future work will be carried out from the following three aspects: 1) enlarge the size of the database; 2) develop region selection models; 3) design algorithms with good performances for SCPs.

\vspace{-0.3cm}
\section{Conclusion}
\vspace{-0.3cm}
In this paper, we focus on a new quality assessment problem about smartphone camera images. First, a new image database SCPQD2020, including 1800 SCPs taken from 120 various scenes by 15 different smartphones, has been built to investigate the subjective quality of SCPs. Each SCP is evaluated from four attributes, which are exposure, color, noise and texture. Moreover, we compare ten successful objective NR IQA models on the SCPQD2020 database. According to the results of performance analysis, we find that no current objective NR model works well, and the quality measures having high correlation with human visual perception are highly needed. This database has been partially released in IEEE ICME2020 Grand Challenge and will be completely and publicly available later.


\bibliographystyle{IEEEbib}
\bibliography{refs}

\end{document}